\documentclass[10pt,aps,pra,twocolumn,showpacs,superscriptaddress]{revtex4-1}
\usepackage[english]{babel} 	
\usepackage[usenames, dvipsnames]{color} 
\usepackage{graphicx} 
\usepackage{bm} 
\usepackage{amsmath} 
\usepackage{amsthm} 
\usepackage{amssymb} 
\usepackage{times} 
\usepackage[pdftex,colorlinks=true,urlcolor= blue,linkcolor=Blue,citecolor=RedViolet]{hyperref}
\usepackage{multirow}
\begin{document}
\title{Energy transport in the presence of entanglement}
\author{A. A. Cifuentes}
\affiliation{Centro de Ci\^encias Naturais e Humanas, Universidade Federal do ABC, 
             09210-170, Santo Andr\'e, S\~ao Paulo, Brazil}
\author{F. L. Semi\~ao}
\affiliation{Centro de Ci\^encias Naturais e Humanas, Universidade Federal do ABC, 
             09210-170, Santo Andr\'e, S\~ao Paulo, Brazil}
\begin{abstract}
In this work, we investigate how the presence of initial entanglement affects energy transport in a network. The network have sites dedicated to incoherent input or output of energy and intermediate control sites where initial entanglement can be established. For short times, we found that initial entanglement in the control sites provides a robust efficiency enhancer for energy transport. For longer times, dephasing considerably damps the quantum correlations, and the advantage of having initial entanglement tends to disappear in favor of the well known mechanism of noise-assisted transport. 
We careful study these two mechanisms in this work, and we believe our findings may be usuful for a better understanding of the relation between nonclassicality and transport, a topic of potential interest for quantum technologies.
\end{abstract}
\maketitle
\section{Introduction}
Quantum control \cite{rabitz2009} and quantum transport (QT) \cite{Mohseni2014} play a prominent role in modern applications of quantum dynamics. In particular, quantum networks (QN) are extensively used to investigate the phenomenon of energy propagation and its relation to quantum coherence. Paradigmatic examples are that of light harvesting complexes \cite{Plenio2008,Caruso2009,Chin2010,Chin2012,Ai2012} and nano devices \cite{Semiao2010}. This kind of investigation led to the discovery of noise-assisted transport (NAT) \cite{Plenio2008,mohseni}, where the interaction with the environment helps to enhance transport efficiency when the system parameters are appropriately tuned. In the context of light harvesting complexes, there are investigations about the capability of the dynamics in promoting the manifestation of quantum correlations \cite{epower,epower2,corr}. Unfortunately,  complete control over system state preparation and its evolution is not yet possible in real photosynthetic systems, let alone the use of characterization tools from quantum information science.

A natural step is then to ask how especially arranged QN of \textit{controlled} quantum systems can be used to critically assess the role of quantum coherences and correlations in the phenomenon of energy transport \cite{Boiuna,Boiuna2}. In this work, we are interested in this kind of investigation. In other words, we are interested in the \textit{active} role of state preparations and external control over transport efficiency. These situate our work in the context of quantum technologies where, instead of having a naturally occurring network, such as a photosynthetic complex, one can deliberately engineer a system, prepare its states, and drive it to study its response. This is precisely the case of setups such as trapped ions, cavity or circuit quantum electrodynamic systems, just to name a few examples. In these systems, control is usually achieved by means of interaction with external fields. In particular, it has been recently shown that cleverly chosen external time dependent drivings can assist quantum tunneling between nodes in a QN \cite{Grifoni1998,Bermudez2011,Bermudez2012}. Here, we employ this ideia to engineer a QN suitable to our purpose of studying how quantum correlations actively influence energy transport.

This article is organized as follows. 
In Section~\ref{model}, we present the network model used in this work. 
Our results are presented in Section~\ref{results}, 
where we carefully investigate the role of quantum correlations and environment in the efficiency of energy transport through the network.
In Section~\ref{conc}, we conclude our findings and present our final remarks. 
\section{Network and Equation of Motion} \label{model}         
\begin{figure}[!h] 
\includegraphics[width=6.5cm, trim = 10 10 0 0]{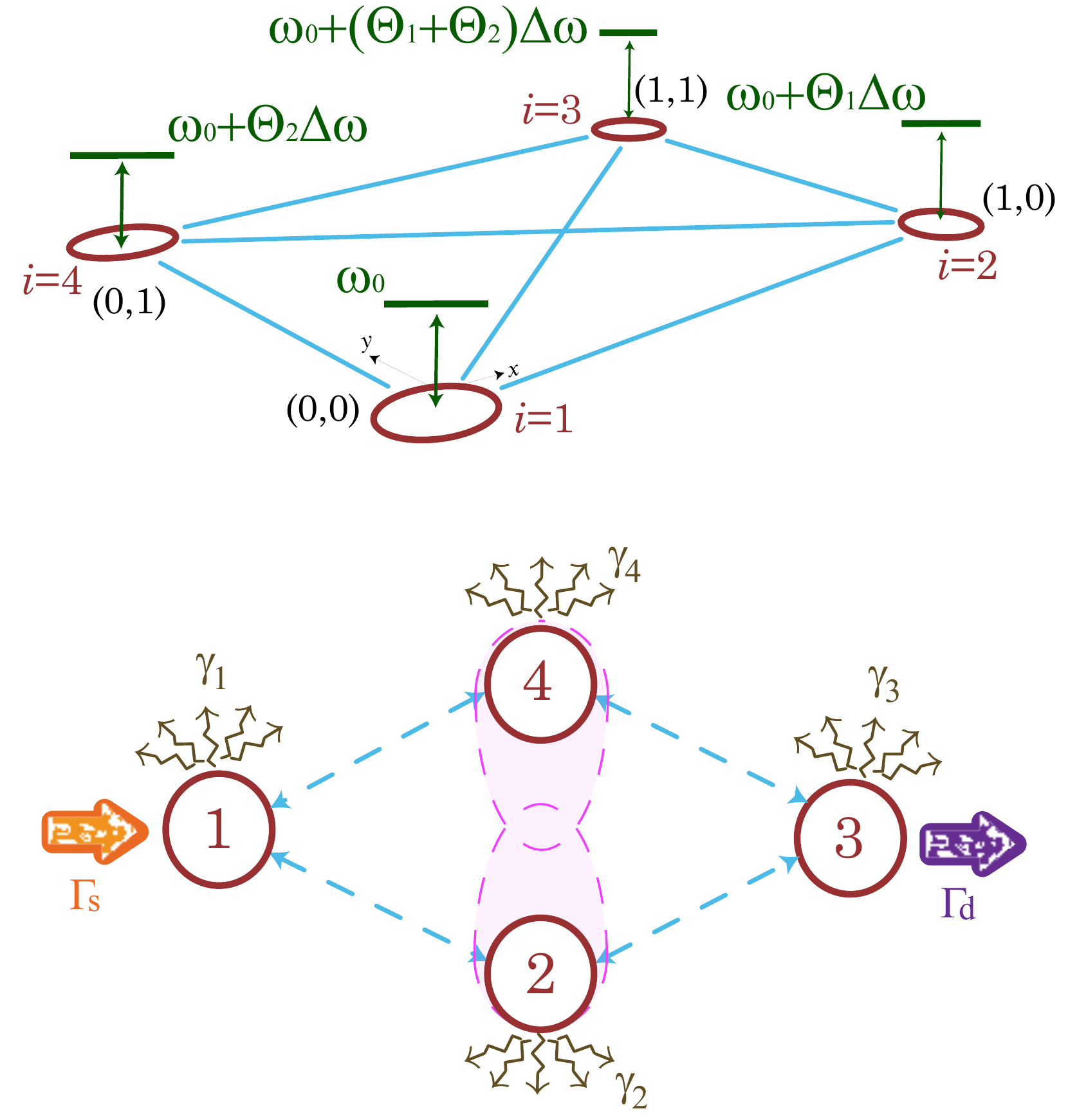} 
\caption{
(Color Online) The network and its basic elements. Top: Four two-level systems labeled by $i$ occupy site ${\bf{i}}=(i_1,i_2)$ forming the network. The lines connecting the sites represent their coherent interaction. The site-dependent energies area also indicated. Bottom: Different dynamical elements of the effective network. It is represented the energy pump ($\Gamma_s$) in site $1$, the energy drain ($\Gamma_d$) in site $3$, the decoherence $\gamma_k$, and the effective energy hopping between sites (two-ended dashed arrows). Also, the possibility of having entangled states initially prepared for the sites $2$ and $4$ is represented by the shadowed area delimited by a dashed contour. 
}                                                                                        \label{fig:Fig1} 
\end{figure}

The basic two-dimensional network element we are interested in is depicted in Fig.\ref{fig:Fig1} (top), where one can see the spatial distribution of the sites or nodes. 

We say that site $i$ ($i=1,2,\ldots$) is located at ${\bf{i}}=(i_1,i_2)$, where $i_1$ and $i_2$ are natural numbers (zero included). By closely following the driving mechanism in  \cite{Bermudez2011,Bermudez2012}, we consider the system Hamiltonian to be
\begin{eqnarray}\label{ht}
\hat{\mathcal{H}}\left(t\right) & = & \hat{\mathcal{H}}_{0}\left(t\right)+\hat{\mathcal{H}}_{c},
\end{eqnarray}
with the total on-site energies subjected to external driving described by 
\begin{equation}\label{h0}
\hat{\mathcal{H}}_{0}\left(t\right) = \hbar \sum_{j}\left[\omega_{j}+\Delta\omega_{j} +\eta_{d,j}\omega_{d,j}\cos\left(\omega_{d,j}t+\phi_{j}\right)\right]\hat{\sigma}_{j}^{+}\hat{\sigma}_{j}^{-},
\end{equation}
and the coherent hopping given by
\begin{eqnarray}\label{hop}
\hat{\mathcal{H}}_{c} & = & \hbar \sum_{ij;k>j}c_{jk}\left(\hat{\sigma}_{j}^{+}\hat{\sigma}_{k}^{-}+\hat{\sigma}_{j}^{-}\hat{\sigma}_{k}^{+}\right),
\end{eqnarray}
where 
\begin{equation}
\Delta\omega_{j}=\Delta\omega\left(\Theta_{1}j_{1}+\Theta_{2}j_{2}\right),
\label{esc}
\end{equation} 
with $\Theta_1, \Theta_2$ positive integers \cite{Gluck2002}, $\hat{\sigma}_{j}^{+}$ and $\hat{\sigma}_{j}^{-}$  two-level raising and lowering operators at site ${{j}}$, respectively,  $\eta_{d,j}$ and $\omega_{d,j}$ are basically the amplitude and angular frequency of the driving field acting on site ${{j}}$, and
 $c_{jk}$ the coherent transfer rate between the sites  ${{j}}$ and  ${{k}}$. More precisely, the product $\eta_{d,j}\omega_{d,j}$ gives the strength of the interaction external drive-site $j$ and it depends monotonically on the field amplitude and frequency. Also, the external drive frequency is chosen to be site-dependent with the phases
 \begin{eqnarray}\label{phase}
 \phi_{{i}}=i_1\phi_x+i_2\phi_y.
 \end{eqnarray}
All these driving parameters and the energy ladder $\Delta\omega_{j}$ are controlled externally and provide a variable tool to design effective interactions in the network \cite{Bermudez2011,Bermudez2012}. 

For the sake of simplicity, we will now choose the driving field controlled parameters such that  $\omega_{d,j}=\omega_d$ and $\eta_{d,j}=\eta_d$, for each site $j$. For the network depicted in Fig.\ref{fig:Fig1} (top), we chose initially resonant sites $\omega_j=\omega_0$ and  $\Delta\omega=r\omega_{d}$,  with $r$  a positive integer, which forms an energy ladder implemented via the time-independent part of the external field. In particular, we depicted the case $r=1$ in Fig.\ref{fig:Fig1} (top). From Eq.(\ref{esc}), it follows that $\Delta\omega_1=0,\,\Delta\omega_2=\Theta_1\Delta\omega,\,\Delta\omega_3=(\Theta_1+\Theta_2)\Delta\omega$, and $\Delta\omega_4=\Theta_2\Delta\omega$. Also, by using Eq.(\ref{phase}), one finds that 
 \begin{eqnarray}
 \phi_1&=&0,\nonumber\\
  \phi_2&=&\phi_x,\nonumber\\
   \phi_3&=&\phi_x+\phi_y,\nonumber\\ 
   \phi_4&=&\phi_y.
 \end{eqnarray}
 The hopping Hamiltonian (\ref{hop}) naturally appears in varied scenarios. It may represent, for instance, dipole-dipole interaction among two-level atoms or molecules in free space \cite{dip1}. Another well known situation where Hamiltonian (\ref{hop}) appears is the dispersive interaction of two-level systems with a common bosonic mode \cite{dip2}. The role of the driving in Eq.(\ref{h0}) is to suppress the coupling between particular pairs of sites. Our goal is to carefully choose the phases and amplitudes of the external driving field to design an effective regime where the transitions between sites $1\leftrightarrow 3$ and $2\leftrightarrow 4$ are suppressed. With this, we can  study how energy injected in site $1$ arrive at site $3$ through indirect pathways $1\leftrightarrow 2\leftrightarrow 3$ and $1\leftrightarrow 4\leftrightarrow 3$, as depicted in Fig.\ref{fig:Fig1} (bottom). This is a situation found, for instance, in the description of the conduction of potassium ions in the KcsA  channel \cite{Cabral,Berne,Cifuentes}. This suppression is achieved as follows.

By transforming the system Hamiltonian (\ref{ht}) to an interaction picture with respect to $\hat{\mathcal{H}}_{0}$, and taking into account the condition $\Delta\omega=r\omega_{d}$ ($r$  a positive integer),  a rotating wave approximation (RWA) can be performed to obtain
%
\begin{eqnarray}
\label{Hint}
\hat{\mathcal{H}}_{I} & = & \hbar \sum_{ij;k>j} \tau_{jk}\left(\hat{\sigma}_{j}^{+}\hat{\sigma}_{k}^{-}+\hat{\sigma}_{j}^{-}\hat{\sigma}_{k}^{+}\right), 
\end{eqnarray} 
with
\begin{eqnarray}\label{aux1}
\tau_{jk}\equiv c_{jk}\mathcal{F}_{f(r,j,k)}\left(\eta_{d},\Delta\phi_{j,k}\right)\text{e}^{-i\,\frac{f(r,j,k)}{2}\left(\phi_{j}+\phi_{k}\right)},
\end{eqnarray}        
where 
\begin{eqnarray}
f(r,j,k) \equiv r\left[\left(\Theta_{1}j_{1}+\Theta_{2}j_{2}\right)-\left(\Theta_{1}k_{1}+\Theta_{2}k_{2}\right)\right],
\end{eqnarray}                                                           
\begin{eqnarray}\label{delta}
\Delta\phi_{j,k} \equiv \phi_{j}-\phi_{k},
\end{eqnarray}
and
\begin{equation}\label{Ec:funcF}
\mathcal{F}_{\chi}\left(\xi,\zeta,\theta\right) \equiv \sum_{s=-\infty}^{\infty}\mathcal{J}_{s}\left(\xi\right)\mathcal{J}_{s+\chi}\left(\zeta\right)\text{e}^{i\,\left(s+\frac{\chi}{2}\right)\theta},
\end{equation}
with $\mathcal{J}_{s}$ being the Bessel function of first kind and order $s$. The rotating wave approximation used to obtain (\ref{Hint}) is valid only for $c_{j,\,k}\ll\omega_0,\omega_{d}$ \cite{Boiuna2}. 

\begin{figure}[!h] 
\includegraphics[width=8.5cm, trim = 10 10 0 0]{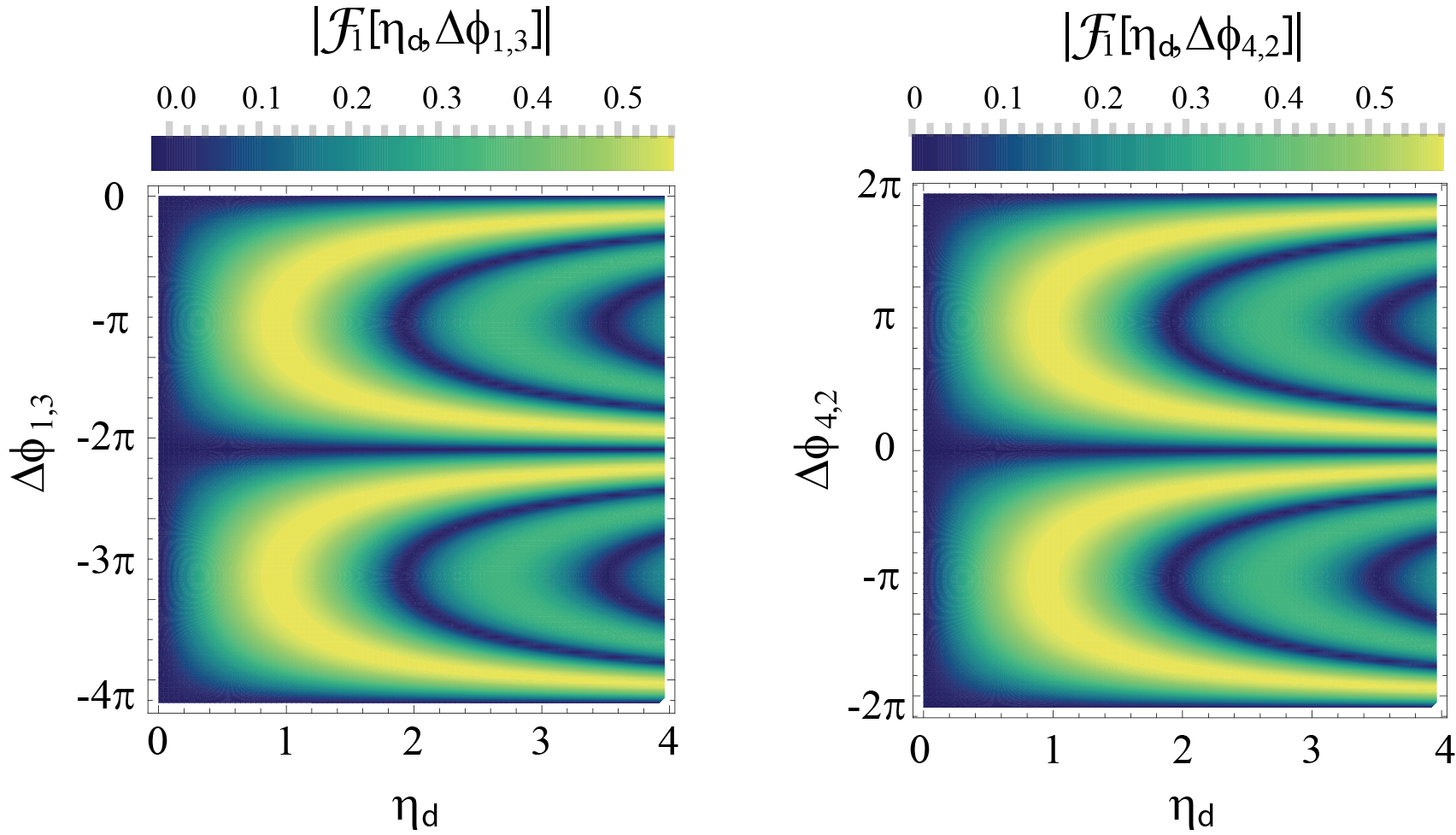} 
\caption{
(Color Online) Hopping amplitude between the sites $j$ and $k$ [Eq.~(\ref{Ec:funcF})] as a function of the driving amplitude $\eta_d$ and  phase $\Delta\phi_{jk}$.  In particular, it is considered $\left|\mathcal{F}_1\right|\equiv\left|\mathcal{F}_{f(1,j,k)}\right|$ where $r=1$.  Left: Sites $\text{{1}}$ and $\text{{3}}$. Right: Sites $\text{{4}}$ and $\text{{2}}$. The parameters used in this plots were $\Theta_1=1$ and $\Theta_2=0$.}                                                                                        \label{fig:Fig2} 
\end{figure}

Now, we carefully look into the content of Eq.~(\ref{Hint}).  For the sake of simplicity, let us consider once again $r=1$. The dynamics of energy migration between sites $i$ and $j$ is ruled by $\mathcal{F}_{f(1,j,k)}$ (see Eq.(\ref{aux1})). The magnitude of this quantity for the pair of diagonal sites $1$-$3$ and $2$-$4$ is plotted in Fig.~\ref{fig:Fig2}  as a function of the driving parameters.  

Direct inspection of plots in Fig~\ref{fig:Fig2} and the use of Eq.(\ref{delta}) thus reveal that the choice $\phi_x=\phi_y=\pi$  leads to the sought suppression of hopping along those diagonal sites. This is true regardless of the driving amplitude $\eta_d$ for the range of parameters considered. It is this choice of $\phi_x$ and $\phi_y$ that will be used from now on. We must now emphasize, however, that we will still keep using the full original (non RWA) time-dependent Hamiltonian (\ref{ht}) in the simulations. The RWA argument was just used to understand how the external driving can effectively suppress hopping between particular pair of sites.

The dynamics of the QN depicted in Fig.\ref{fig:Fig1} (bottom), which is our system of interest, is ruled by the following master equation  \cite{Carmichael2002}
\begin{equation}
\label{eq:master_eq}
\frac{d\hat{\rho}}{dt} = -\frac{i}{\hbar}\left[\hat{\mathcal{H}}\left(t\right),\hat{\rho}\right]+\mathcal{L}_{s}(\hat{\rho})+\mathcal{L}_{d}(\hat{\rho})+\mathcal{L}_{deph}(\hat{\rho}),
\end{equation}
where the {\it{source}} superoperator
\begin{equation}\label{s}
\mathcal{L}_{s}(\hat{\rho}) = \Gamma_{s}\left(-\left\{ \hat{\sigma}_{1}^{-}\hat{\sigma}_{1}^{+},\hat{\rho}\right\} +2\hat{\sigma}_{1}^{+}\hat{\rho}\hat{\sigma}_{1}^{-}\right)
\end{equation}
accounts for the incoherent input of energy into the system through site $1$ at a pump rate $\Gamma_{s}$, the {\it{drain}} superoperator
%
\begin{equation}\label{d}
\mathcal{L}_{d}(\hat{\rho}) = \Gamma_{d}\left(-\left\{ \hat{\sigma}_{3}^{+}\hat{\sigma}_{3}^{-},\hat{\rho}\right\} +2\hat{\sigma}_{3}^{-}\hat{\rho}\hat{\sigma}_{3}^{+}\right)
\end{equation}
represents an incoherent loss of energy through site $3$ at a rate $\Gamma_{d}$, and the dephasing superoperator
%
\begin{equation}
\mathcal{L}_{deph}(\hat{\rho}) = \sum_{k=1}^{N}\gamma_{k}\left(-\left\{\hat{\sigma}_{k}^{+}\hat{\sigma}_{k}^{-},\hat{\rho}\right\} +2\hat{\sigma}_{k}^{+}\hat{\sigma}_{k}^{-}\hat{\rho}\hat{\sigma}_{k}^{+}\hat{\sigma}_{k}^{-}\right),
\end{equation}
destroys quantum coherence in the network, where $\gamma_{k}$ is a site-dependent dephasing rate.  
In all these,  $\left\{\star,\rho\right\}$ denotes the anticommutator $\left\{\star,\rho\right\}\equiv\star\,\rho+\rho\,\star$. One of our goals is to study the contribution of each of these terms in the master equation to the dynamics of energy propagation in the QN depicted in Fig.\ref{fig:Fig1} (bottom). Moreover, we want to investigate it taking into account the presence of initial quantum correlations in the control sites $2$ and $4$.

An important figure of merit is how efficiently energy leaves the system through site $3$ which works as a drain (rate $\Gamma_d$). This is quantified by the  integrated population of site $3$ 
\begin{equation}\label{ind}
P_{3} = \int_{0}^{t}p_{33}(t')dt', 
\end{equation}
where $p_{33}(t')$ is the occupation of site $3$ at instant $t'>0$. This quantity is proportional to the  {\it{transport efficiency}} $\eta_{\text{eff}}$ through $P_{3}=\eta_{\text{eff}}/2\Gamma_d$ \cite{mohseni,Rebentrost2009}. 

In this work, we use the entanglement of formation (EoF) to quantify bipartite entanglement between pairs of sites  \cite{Wootters1998}. In addition to that, we also include other form of quantum correlation in our study, the so called Quantum Discord (QD) \cite{Ollivier2001}. The former is interesting because it spots quantumness for a set of states which does not necessarily contains entanglement. In this sense, the QD adds generality to our study.
\section{Results} \label{results}      
The just presented formalism allows one to explore site-dependent dephasing scenarios. However, from now on we will be adopting the same decoherence rates for all sites $\gamma_k=\gamma$, a feasible choice for quantum technologies. Nevertheless, if one aims at studying transport in natural systems such as photosynthetic complexes, site-dependent dephasings should necessarily be taken into account in accordance with experimental observations and computational simulations.                     
\subsection{Efficiency enhancers - entanglement versus dephasing} \label{sect_pad}                         
We would like to start our investigations by considering two initial states, with same mean energy (one excitation),  but with different types of correlations. The first state is 
\begin{equation}\label{eq:rho0_init_ent} 
\rho_{ent}=\left|\psi\right\rangle\! \left\langle \psi\right|,
\end{equation}
where $\left|\psi\right\rangle=\left(\left|gegg\right\rangle +\left|ggge\right\rangle\right)/ \sqrt{2}$. For this initial preparation, sites $1$ and $3$ start in their ground state and they are not correlated with sites $2$ and $4$, which share a bipartite maximally entangled state that contains one quantum of excitation. Experimentally, bipartite entangled states such as the one considered here has been generated in a variety of setups ranging from photons \cite{Bell1} to massive particles \cite{Bell2}. In the scenario defined by Eq.(\ref{eq:rho0_init_ent}), sites $2$ and $4$ are then quantum correlated. The second initial state to be considered in this subsection is 
\begin{equation}\label{eq:rho0_init_mix} 
\rho_{mix}=\left(\left|gegg\right\rangle \left\langle gegg\right|+\left|ggge\right\rangle \left\langle ggge\right|\right)/2,
\end{equation}
where once again the sites $1$ and $3$ are initially in their ground state, but now sites $2$ and $4$ are just classically correlated in a maximally mixed state. One can consider Eq.(\ref{eq:rho0_init_mix}) as the limit of Eq.(\ref{eq:rho0_init_ent}) when previous decoherence $(t<0)$ on the decoupled model $(c=0)$ had fully acted and completely destroyed the coherences i.e., the nondiagonal terms in the basis $\{|gegg\rangle, |ggge\rangle\}$.

\begin{figure}[!h] 
\includegraphics[width=7.5cm, trim = 10 10 0 0]{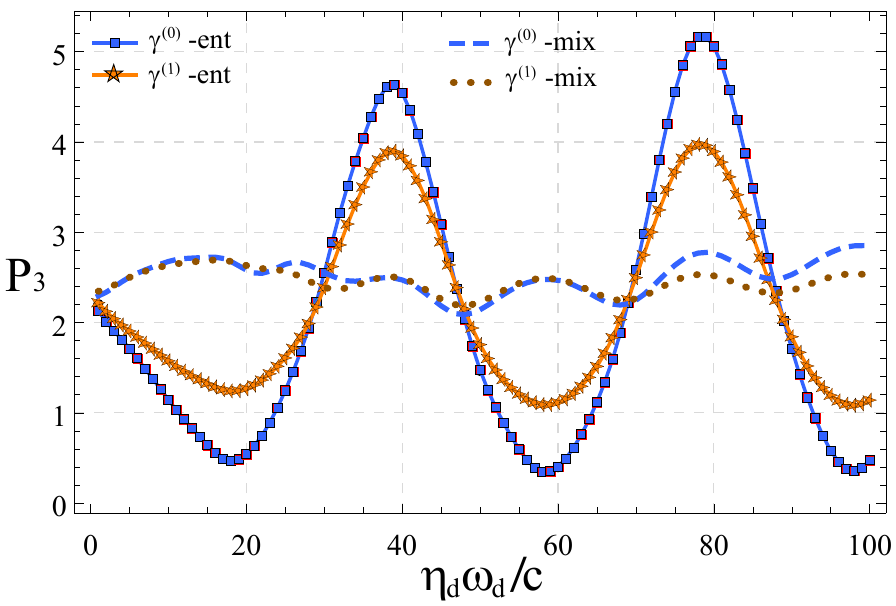} 
\caption{
(Color Online) Integrated population of site $3$ [Eq.(\ref{ind})] as a function of the \textit{rescaled} driving strength $\eta_d\omega_d/c$, for the choices: $\omega_j=\omega_0\,\,\forall\,j$, $\Delta\omega=\omega_0/4$. $\phi_x=\pi$, $\phi_y=\pi$. $\Theta_1=1,\,\,\Theta_2=0$ (energy ladder along $x$ direction). $c_{j,k}=c=\omega_0/100\,\,\forall j,k$, $\omega_d=\omega_0/4$. $\gamma_k=\gamma\,\,\forall k$, $\gamma^{(0)} =\gamma= 0$, $\gamma^{(1)}=\gamma= c/10$. $\Gamma_d=c/100$,  $\Gamma_s=2\,\Gamma_d$. {\bf{-ent}}: Initial state is Eq. (\ref{eq:rho0_init_ent}).  {\bf{-mix}}: Initial state is Eq. (\ref{eq:rho0_init_mix}).}                                                     \label{fig:Fig3} 
\end{figure}

In Fig.~\ref{fig:Fig3}, we present the efficiency quantifier (\ref{ind}) for an observation time $ct=10$ (interval of integration) and a dephasing rate $\gamma=c/10$ what means that the dephasing times are around one order of magnitude longer than the hopping times (coherent dynamics). Within this time interval, which can be called a short interaction time, quantum effects have a chance to manifest or to bring some influence on transport. In the long-time regime, to be briefly discussed next, quantum effects usually become irrelevant since dephasing generally kills the coherences. 

In this scenario, by comparing the plots  in Fig.~\ref{fig:Fig3}, it is clear that the initial presence of entanglement helped transport, i.e., resulted in values of $P_3$ that surpassed those obtained with the initially mixed (non-entangled) situation.  In other words, entanglement worked as a {\it{resource}} for QT. Another interesting feature of Fig.~\ref{fig:Fig3} is the presence of NAT for the minima when entanglement was originally present in sites $2$ e $4$. By increasing the dephasing, the efficiency increased in those regions. This is not true for the maxima where dephasing tends to be destructive. This is expected because dephasing destroys entanglement in our model, and entanglement is precisely the ingredient for the pronounced maxima. For the initial mixed state, the phenomenon of NAT is practically absent in Fig.~\ref{fig:Fig3}. 

\begin{figure}[!htbp] 
\includegraphics[width=8.5cm, trim = 10 10 0 0]{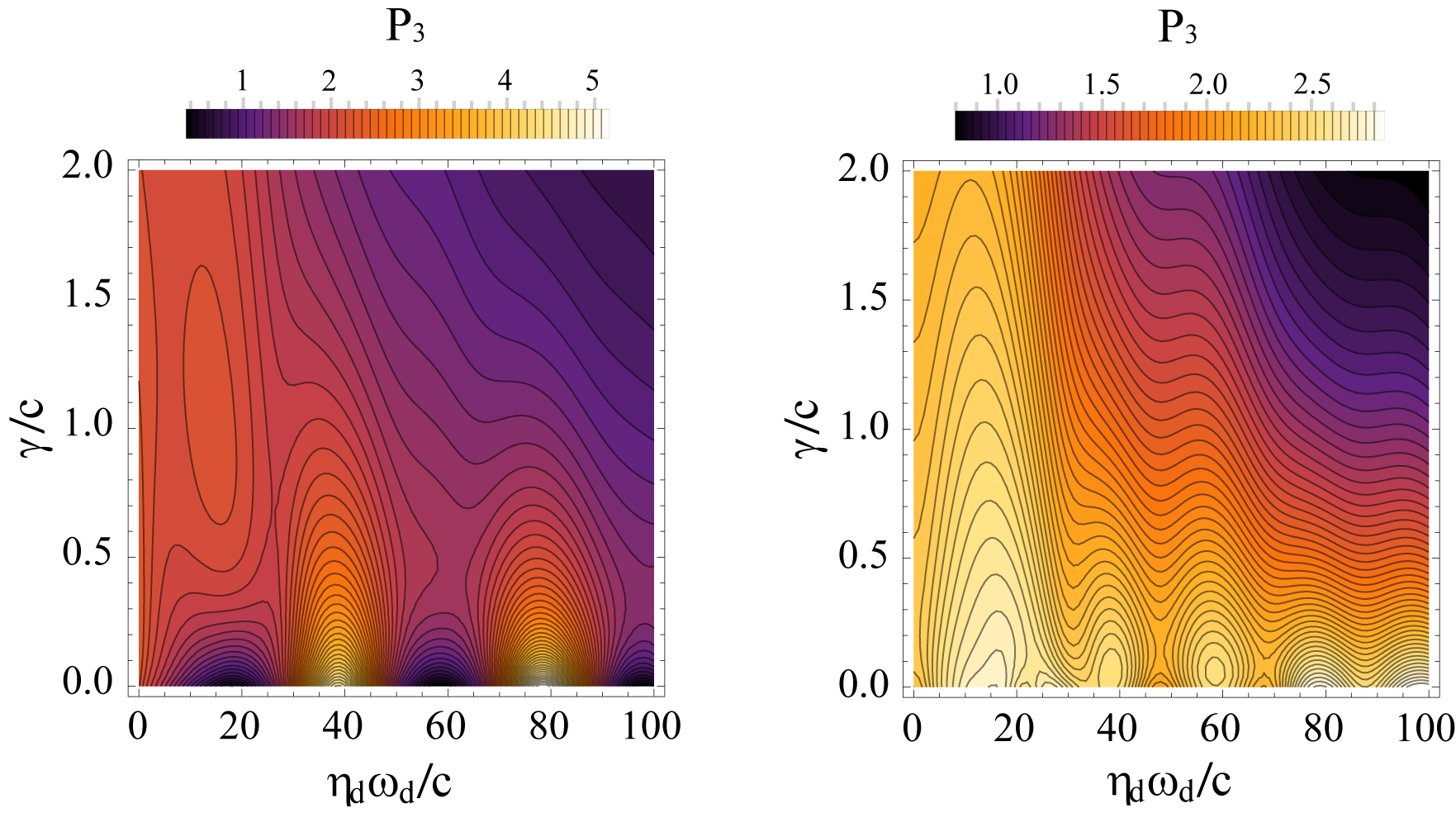} 
\caption{
(Color Online) Efficiency indicator $P_3$ as a function of the \textit{rescaled} driving strength $\eta_d\omega_d/c$, and the dephasing $\gamma/c$. The parameters for the plot are as in Fig.~\ref{fig:Fig3}. Left: Considering the initial state Eq.  (\ref{eq:rho0_init_ent}). Right: Considering the initial state Eq.  (\ref{eq:rho0_init_mix}).}                                                     \label{fig:Fig3_1} 
\end{figure}

All these interesting features are confirmed by Fig.~\ref{fig:Fig3_1} where we provide a more general picture of the problem through variation of dephasing over a broad range. It is interesting to see that for the initial preparation with entanglement (plot on the left),  increasing the dephasing $\gamma$ is generally beneficial for the minima, and that this feature is practically not manifested for the non-entangled initial situation (plot on the right). On the contrary, dephasing acted as a hinderer when no entanglement was initially present.
%
%

\begin{figure}[!htb] 
\includegraphics[width=7.5cm, trim = 0 5 0 0]{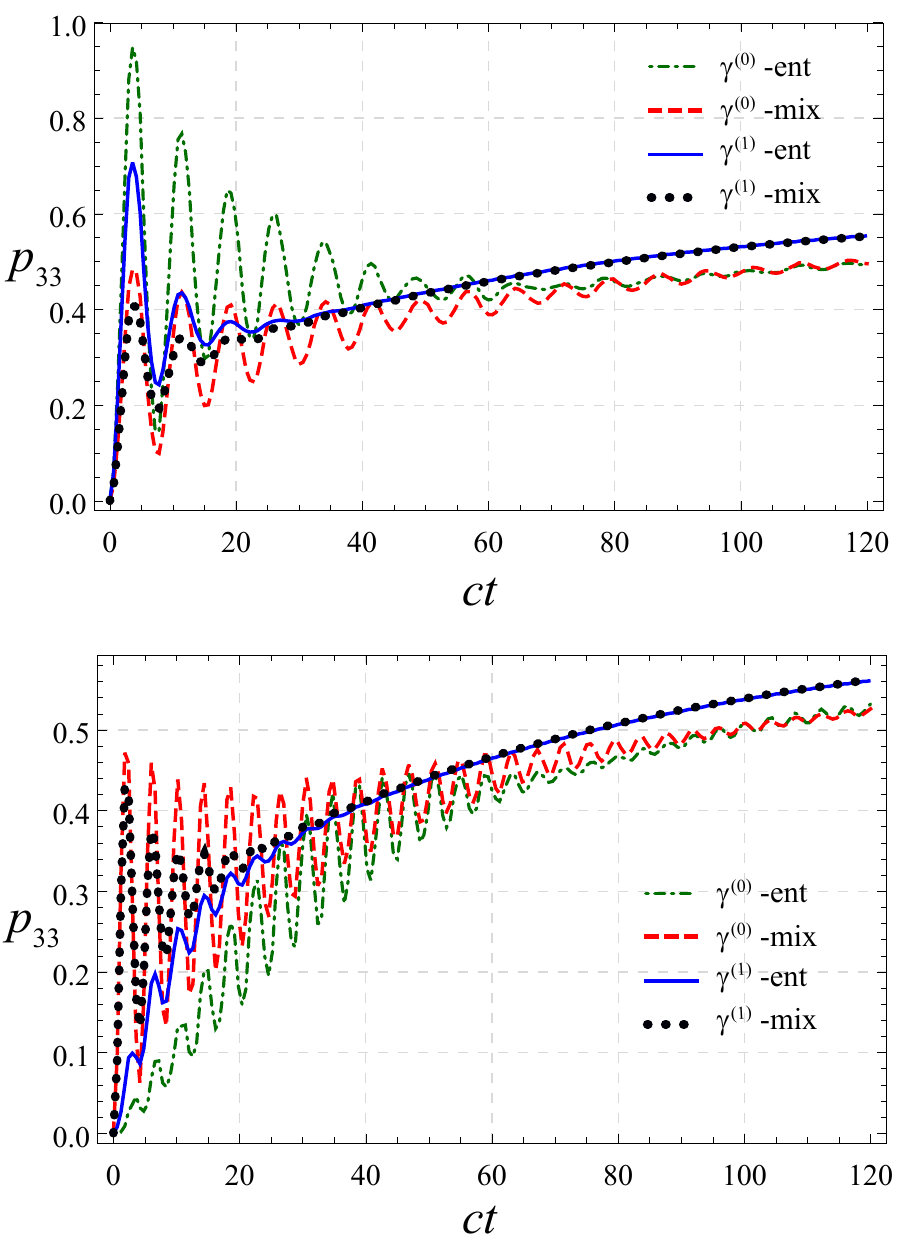} 
\caption{
(Color Online) Temporal dependence of the population of site $3$, $p_{33}$, for the first minimum ($\eta_d\omega_d/c=18.0$) and maximum ($\eta_d\omega_d/c=38.8$) values in Fig.~\ref{fig:Fig3}. {\bf{-ent}}: Initial state is Eq. (\ref{eq:rho0_init_ent}).  {\bf{-mix}}: Initial state is Eq. (\ref{eq:rho0_init_mix}). Top: First maximum. Bottom: First minimum. 
The other parameters for the plot are as in Fig.~\ref{fig:Fig3}.
}                                                                                        \label{fig:Fig4} 
\end{figure}

We now present the long-time behavior of the population $p_{33}$ of the last site, fixing two values of $\eta_d\omega_d/c$:  the first minimum and the first maximum of $P_3$ with initial entanglement in Fig.~\ref{fig:Fig3}. The results are shown in Fig.~\ref{fig:Fig4}, where one can see that initial entanglement, as expected, gradually looses its capacity to boost transport. The only enhancer left is dephasing through NAT. This mechanism is clearly manifest in Fig.~\ref{fig:Fig4} giving the fact that, for times longer than $ct\approx 60$, the curves with non null dephasing $\gamma^{(1)}$ are above the ones with null dephasing $\gamma^{(0)}$.

\begin{figure} 
\includegraphics[width=7.5cm, trim = 0 5 0 0]{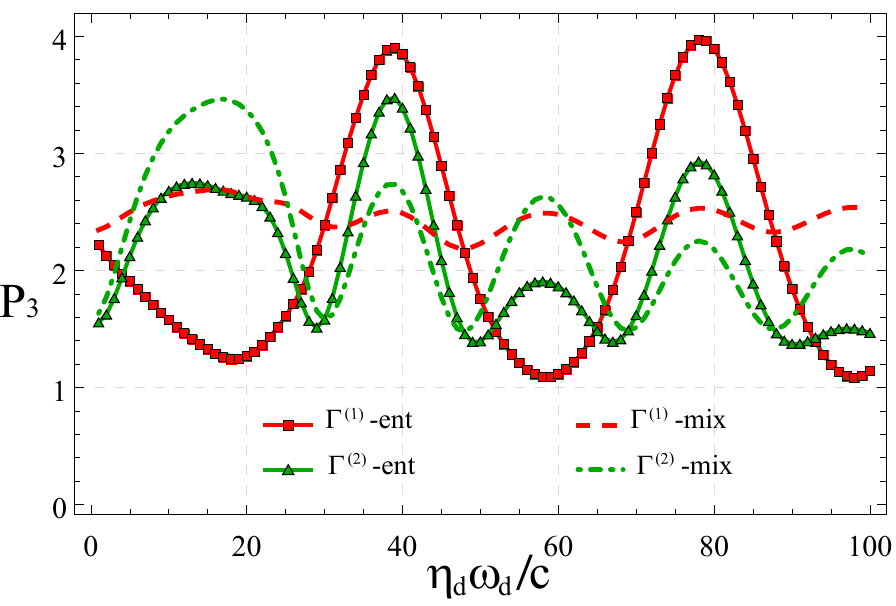} 
\caption{
(Color Online)  Integrated population of site $3$, for the two {\it{drain}} rates $\Gamma^{(1)}=c/100$ and $\Gamma^{(2)}=c/10$. The chosen dephasing rate is $\gamma= c/10$. {\bf{-ent}}: Initial state is Eq. (\ref{eq:rho0_init_ent}).  {\bf{-mix}}: Initial state is Eq. (\ref{eq:rho0_init_mix}).  
The other parameters for the plot are as in Fig.~\ref{fig:Fig3}.}                                                                                        \label{Gamd} 
\end{figure}

Before finishing this section, we would like to go a little deeper in our investigation about the entanglement-assisted transport phenomenon already observed in previous plots. We will do this in two directions. First, we would like to see how Fig.~\ref{fig:Fig3} changes when the input of energy is changed. This is shown in Fig.~\ref{Gamd}. The plots are very illustrative because they once again show the competition between entanglement-assisted transport and noise-assisted transport, now for a different scenario where the increase of noise comes from other sources than pure dephasing. For $\Gamma^{(1)}$, the beneficial effect coming from initial entanglement is still quite clear, almost like in Fig.~\ref{fig:Fig3}. To see this, compare, for instance, the  maxima arising from the situation with initial maximal entanglement with the situation with and maximal mixedness. When the incoherent input of energy is increased to $\Gamma^{(2)}$,  the initial maximally mixed state and the initial maximally entangled state are practically equivalent in terms of efficiency.  The reason why initial entanglement starts loosing importance when the energy input rate increases is that adding more energy also adds more noise. This is so because the energy input and output are both incoherent process corresponding to non-unitary terms in the network master equation, see Eqs.(\ref{s}) and (\ref{d}).

\begin{figure} 
\includegraphics[width=6.0cm, height=6.2cm, trim = 0 5 0 0]{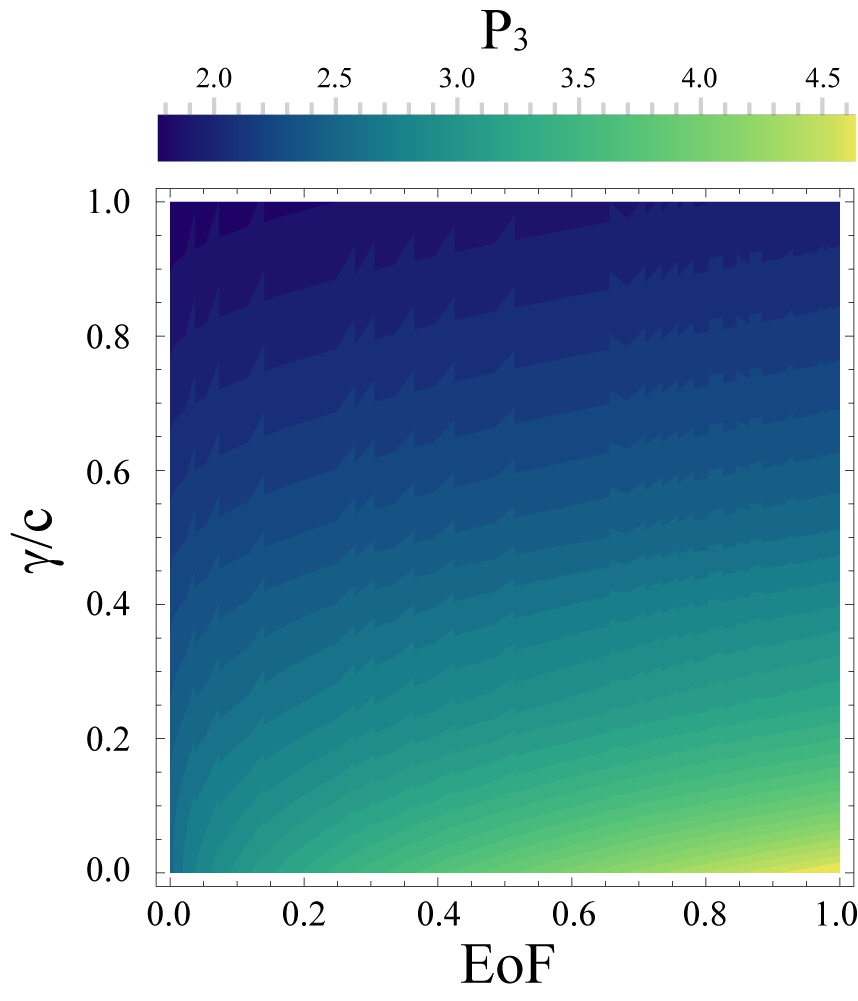} 
\caption{
(Color Online) Efficiency indicator $P_3$, as a function of dephasing and entanglement in the initial state, for the first maximum ($\eta_d\omega_d/c=38.8$) in Fig.~\ref{fig:Fig3}. The {\it{drain}} rate is $\Gamma_{d}=c/100$. The other parameters are as in Fig.~\ref{fig:Fig3_1}.}                                                                                        \label{entvar} 
\end{figure}

The second direction we want to explore is the variation of the initial entanglement. Up to now, we worked only with maximal entanglement versus non-entanglement at all, both states with only one excitation shared between sites $2$ and $4$. We now keep considering the one excitation sector, but with the state $|\psi\rangle=\cos\theta|eg\rangle+\sin\theta|ge\rangle$ for sites $2$ and $4$. The other two sites are still considered to be initially in the ground state. The variation of $\theta$ makes the entanglement of formation vary from zero $(\theta=0)$ to one $(\theta=\pi/4)$. In Fig.~\ref{entvar}, we once again consider the efficiency indicator $P_3$ in the first maximum $\eta_d\omega_d/c=38.8$ [see Fig.~\ref{fig:Fig3}]. However, we now have it as a function of dephasing and the entanglement in the initial state $|\psi\rangle$. One can see that states with more initial entanglement lead to higher values of $P_3$. This confirms the robustness of the entanglement-assisted transport mechanism for a whole class of states (all pure states with one excitation shared by control sites $2$ and $4$). Finally, one can see once again that, for initial pure states and the maxima of $P_3$ in Fig.~\ref{fig:Fig3}, the increase of dephasing $\gamma$ is a hindrance to transport.
\subsection{Quantum correlations survival}
\begin{figure}[!t] 
\includegraphics[width=7.2cm, trim = 0 5 0 0]{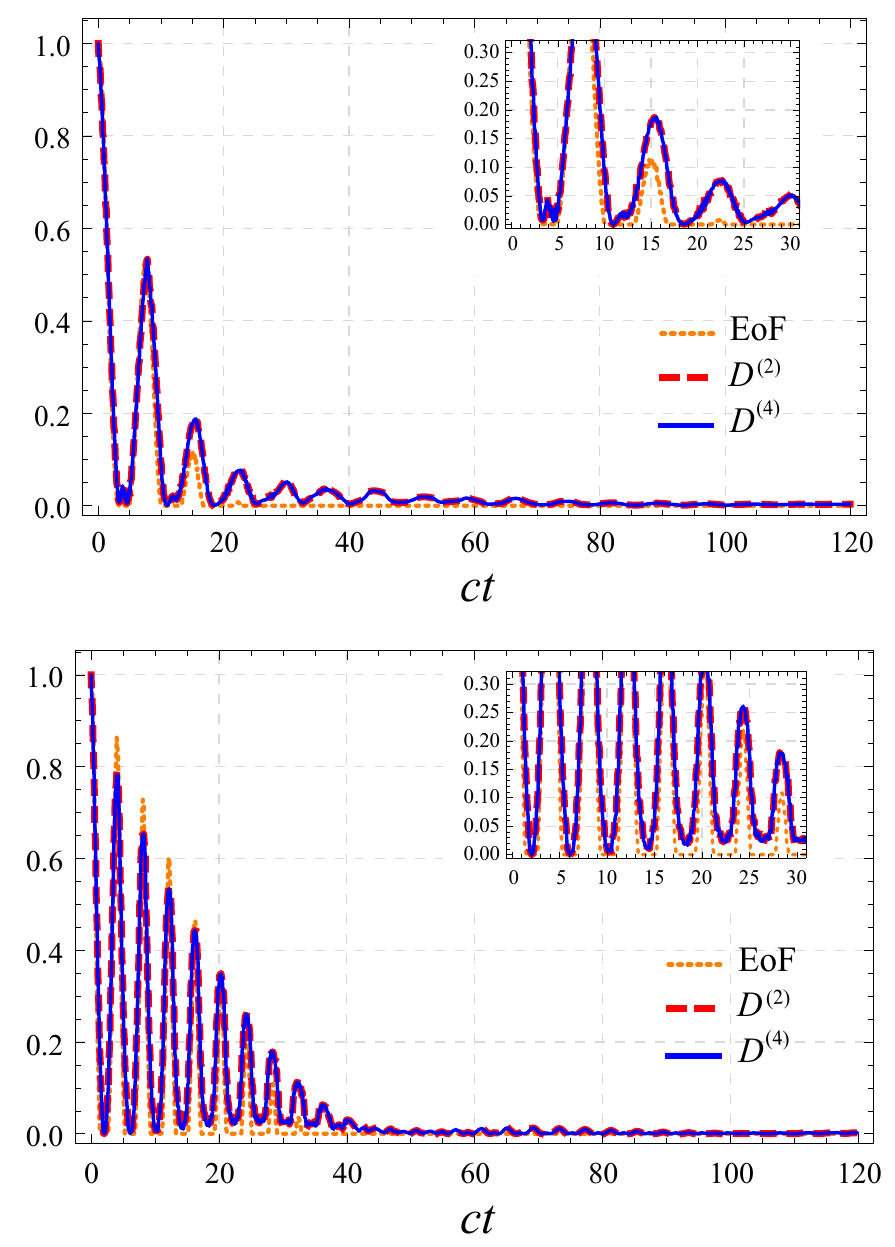} 
\caption{(Color Online) Temporal dependence of the QC between the sites $\text{\bf{4}}$ and $\text{\bf{2}}$. Insets: Zoom intended to highlight the existence of QC apart from entanglement for some times. The initial state is Eq. (\ref{eq:rho0_init_ent}). $\gamma_k=0$ $\forall\,\,k$ and $\Gamma_d=c/100$. Top: First maximum. Bottom: First minimum. 
The other parameters for the plot are as in Fig.~\ref{fig:Fig3}.
}                                                                                        \label{fig:Fig5} 
\end{figure}

\begin{figure}[!b] 
\includegraphics[width=7.5cm, trim = 0 5 0 0]{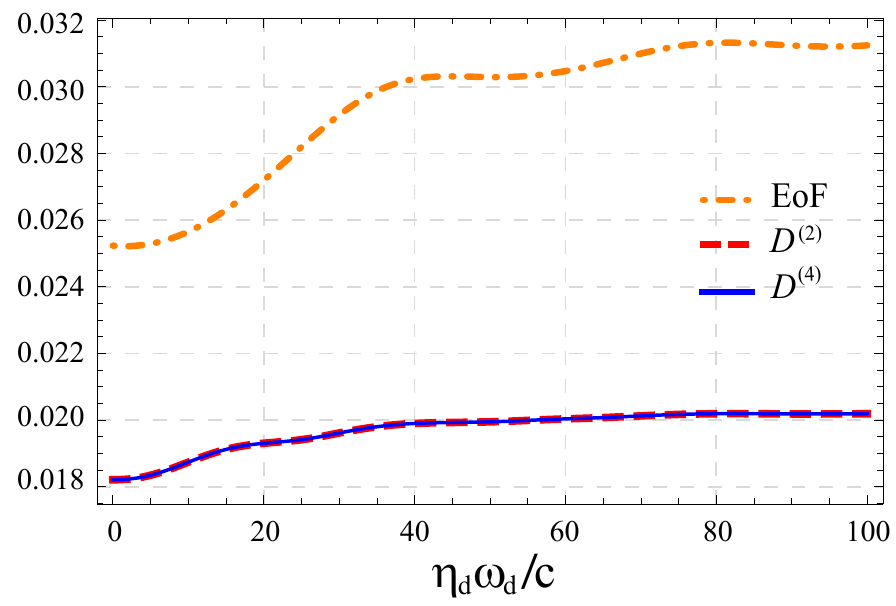} 
\caption{
(Color Online) Temporal average of QC between the sites $\text{\bf{4}}$ and $\text{\bf{2}}$ for the situation where the system is initially in  an entangled state. A large dephasing situation is considered, with $\gamma_k= c$ $\forall\,\,k$ and $\Gamma_d=c/100$, for the initial state Eq. (\ref{eq:rho0_init_ent}). 
The other parameters for the plot are as in Fig.~\ref{fig:Fig3}.
}                                                                                        \label{fig:Fig6} 
\end{figure}

Since the initial presence of entanglement was seen to be beneficial for transport in the short-time behavior, it would be interesting to have a close look at its dynamics. This could help us to better understand the conclusions  previously presented about the effect of quantum correlations (QC) over transport for our system of interest. As said before, we will be employing quantifiers of entanglement and quantum discord QD. For the latter, it is important to distinguish between the situation where projective measurements are thought to act on one subsystem or another. In our case, the subsystems are sites $2$ and $4$, and we will then denote the case where projective measurements are intended to act on $2$ by $D^{(2)}$, and when intended on $4$ by $D^{(4)}$. The dynamical behavior of quantum correlations for the same parameters considered in Fig.~\ref{fig:Fig4} is shown in Fig.~\ref{fig:Fig5}. As expected, at long times the oscillatory behavior of the quantum correlations is completely damped rendering the system state to be essentially classicality correlated. It is interesting to see that there are times where quantum discord remains finite in spite of the fact that entanglement goes to zero. In Fig.~\ref{fig:Fig6}, we present the average correlations over the same time spam consider to evaluate $P_3$ in Fig.~\ref{fig:Fig3}, for an initially maximally entangled situation. From this plot, it is clear that also on average quantum correlations remain finite during the transport for the whole range of driving strengths $\eta_d\omega_d/c$ considered in Fig.~\ref{fig:Fig3}. From Fig.~\ref{fig:Fig6} it is also possible to see that entanglement is a bit more sensitive to the choice of the driving strength than discord in the sense that the former oscillates more stronger than the latter as the driving strength is varied.

\section{Conclusions} \label{conc}                                    
In this work, we studied the relation between transport efficiency and initial presence of entanglement in a network. The network used in this work is the simplest one needed to assess the effect of having entanglement in the intermediate sites, i.e., between two sites not directly connected to energy sources or sinks. In fact, we found that initial entanglement provides a robust enhancer of transport efficiency.  For short times, entanglement-assisted transport showed up for all possible initial pure states with only one excitation in the network. In this time domain, we showed that quantum correlations survive dynamically and on average thus rendering the transport to be quantum in essence. On the other hand, for longer times, these correlations progressively vanish and it comes to a point where only noise-assisted transport is available as a transport enhancer. We also showed that, for short-times, noise-assisted transport is imaterial for the maximally mixed initial situation.  
\acknowledgments 
A.A.C. acknowledges to 
``Coordena\c{c}\~ao de Aperfei\c coa\-men\-to de Pessoal de N\'ivel Superior'' (CAPES).
FLS acknowledges partial support from the Brazilian National Institute of Science and Technology of 
Quantum Information (INCT-IQ) and CNPq 
under Grant No. 307774/2014-7. We would also like to thank Ms. Marcela Herrera and Prof. Roberto Serra for granting access to their computing facilities.

\end{document}